\documentclass[aps,prb,twocolumn]{revtex4}
\usepackage{amsfonts}
\usepackage{amsmath}
\usepackage{amssymb}
\usepackage[final,dvips]{graphicx}
\begin{document}
\title{Slowing down Josephson vortex lattice in
Bi$_{2}$Sr$_{2}$CaCu$_{2}$O$_{8+\delta}$ with pancake vortices}
\author{A. E. Koshelev}
\affiliation{Materials Science Division, Argonne National Laboratory, Argonne, Illinois 60439}
\author{Yu. I. Latyshev}
\affiliation{Institute of Radio-Engineering and Electronics, Russian
Academy of Sciences, Mokhovaya 11-7, 101999 Moscow, Russia}
\author{M. Konczykowski}
\affiliation{Laboratoire des Solides Irradi\'{e}s, CNRS UMR 7642, Ecole Polytechnique,
91128 Palaiseau, France}

\begin{abstract}
We study theoretically and experimentally influence of pancake
vortices on motion of the Josephson vortex lattice in layered
high-temperature superconductors. Mobility of the Josephson vortices
in layered superconductors is strongly suppressed by small amount of
pancake-vortex stacks. Moving Josephson vortex lattice forces
oscillating zigzag deformation of the pancake-vortex stacks
contributing to damping. The salient feature of this contribution is
its nonmonotonic dependence on the lattice velocity and the
corresponding voltage. Maximum pancake effect is realized when the
Josephson frequency matches the relaxation frequency of the stacks.
The pancake-vortex damping is strongly suppressed by thermal
fluctuations of the pancake vortices. This theoretical picture was
qualitatively confirmed by experiments on two mesas prepared out of
Bi$_{2}$Sr$_{2} $CaCu$_{2} $O$_{8+\delta}$ whiskers. We found that
the Josephson-vortex flux-flow voltage is very sensitive to small
c-axis magnetic field. The pancake-vortex contribution to the
current indeed nonmonotonically depends on voltage and decreases
with increasing temperature and in-plane magnetic field. We also
found that irradiation with heavy ions has no noticeable direct
influence on motion of the Josephson vortices but dramatically
reduces the pancake-vortex contribution to the damping of the
Josephson vortex lattice at low temperatures.
\end{abstract}
\date{\today}
\maketitle


\section{Introduction}

The layered crystalline structure of the cuprate high-temperature
superconductors leads to existence of two types of vortices in these
materials, pancake-vortex (PV) stacks \cite{pancakes} induced by the
magnetic field component perpendicular to the layers and Josephson
vortices (JVs) \cite{BulClemPRB91} created by the magnetic field component
parallel to the layers. Repulsive interaction between the vortices of each
type results in formation of regular vortex lattices. In particular, the
magnetic field applied along the layers generates triangular lattice of
the JVs stretched along the layer direction. The anisotropy factor
$\gamma$ sets the important field scale, $B_{\mathrm{cr}
}=\Phi_{0}/(2\pi\gamma s^{2})$ , where $s$ is the interlayer periodicity.
When the magnetic field exceeds $B_{\mathrm{cr}}$ the Josephson vortices
homogeneously fill all layers forming a dense lattice \cite{BulClemPRB91}.
In highly anisotropic materials, like
Bi$_{2}$Sr$_{2}$CaCu$_{2}$O$_{8+\delta}$ (BSCCO) this field scale is
rather moderate $\sim$ 0.5 tesla.

In BSCCO, due to a very weak Josephson interlayer coupling, two
types of vortices can coexist in the tilted magnetic field
\cite{BulLedvKoganPRB92}. The static attractive interaction between
JVs and PV stacks \cite{CrossLatLet} leads to many exotic vortex
states, such as mixed chain-lattice state
\cite{Bolle91,Grig95,MatsudaSci02} and pancake-chains
state\cite{GrigNat01}, and has been used for visualization of JV
stacks \cite{MatsudaSci02,VlaskoPRB02,TokunagaPRB02}, see recent
review \cite{ChainReview}.

\begin{figure}[ptb]
\begin{center}
\includegraphics[width=2.3in]{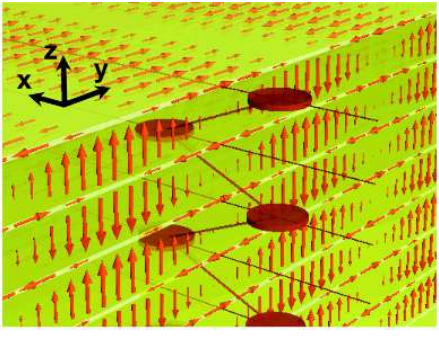}
\end{center}
\caption{Sketch of a pancake-vortex stack in presence of the dense
JV lattice. The arrows illustrate current distribution of the JV
lattice and the discs illustrate pancake vortices. The JV lattice
induces zigzag deformation of the stack. Moving lattice induces
oscillating zigzag deformations which contribute to dissipation, see
animation
.} \label{Fig:JVL-PancStack}
\end{figure}

Dynamic properties of the crossing-lattices state have been studied
in much less details. A particularly interesting topic is dynamic
behavior of the JV lattice. An external transport current flowing
across the layers drives the JV lattice along the layers. Due to
relatively weak intrinsic dissipation, the Josephson vortices can be
accelerated up to very high velocities. Dynamics of the JV lattice
in BSCCO have been extensively studied by several experimental
groups (see, e.g., Refs.\ \onlinecite{Lee,Hechtfischer,Latyshev}).
When magnetic field is tilted at small angle with respect to the
layers, the c-axis field component generates small concentration of
PV stacks. Alternating in-plane supercurrents of static JV lattice
force zigzag deformations of the PV stacks \cite{Bul-zigzag}, see
Fig.\ \ref{Fig:JVL-PancStack}. It is well known that mobility of JVs
is strongly suppressed by a very small amount of PV stacks
\cite{LatPhysC91,EnriquezPRB96,HechtfischerPRB97}. As a consequence,
studying the JV lattice transport always requires a very accurate
alignment of the magnetic field with the layers. In spite of that
common knowledge, JV lattice dynamics in presence of the PV stacks
has never been investigated systematically.

In the case of strong anisotropy, the JV lattice can move through
static PV stacks. Even in this case the PV stacks will strongly
impede motion of the JV lattice. Dynamic behavior of the PV stack
naturally follows from its static configuration. The moving JV
lattice forces oscillations of the PV stacks leading to additional
dissipation and slowing down the lattice. In this paper we
investigate this effect quantitatively in the case of dense JV
lattice. Influence of the PV stacks on motion of an \emph{isolated}
JV has been considered theoretically in Ref.\
\onlinecite{JVPancPRB03}.

The paper is organized as follows. In Sec.\ \ref{Sec:Theory} we
present theoretical models describing influence of the PV stacks on
motion of the dense JV lattice. We compute the dynamic friction
force generated by PV stacks and study suppression of this force by
PV fluctuations. We also consider influence of the PV fluctuations
on the regular JV flux-flow resistivity and influence of columnar
defects on the PV-induced damping of the JV lattice. In Sec.\
\ref{Sec:Experiment} we present experimental results. Studying the
flux-flow of the JV lattice for small c-axis magnetic fields, we
extracted the PV contribution to the JV damping and studied its
dependence on the voltage, temperature, and in-plane field. We also
found that this PV contribution is strongly suppressed by heavy-ion
irradiation. In Sec. \ref{Sec:Discussion} we discuss comparison
between the experiment and theory and possible applications of the
studied effect.

\section{Theoretical models \label{Sec:Theory}}

\subsection{Basic model}

A general description of JV dynamics in presence of the PV stacks is
rather complicated. We consider first the simplest situation, in
which influence of the pancake vortices on motion of the JV lattice
can be traced in a clearest way. We assume that a strong magnetic
field, $B\gtrsim1$ tesla, is applied in the $x$-$z$ plane at a very
small angle with respect to the layers. Such a magnetic field
generates the dense JV lattice and a dilute array of the PV stacks.
We neglect both pinning and thermal fluctuations of the pancake
vortices. We start with derivation of the interaction between the JV
lattice and deformed PV stack via the interlayer Josephson energy,
$\mathcal{E} _{J}=-E_{J}\sum_{n}\int
d^{2}\mathbf{r}\cos{\theta_{n}}$ where $\theta_{n}$ is the phase
difference between the layers $n$ and $n+1$. The static dense JV
lattice is characterized by the phase difference,
$\theta_{Jn}(y)\approx k_H(y-y_{J})+\pi n$ with $k_H=2\pi
sB_{x}/\Phi_{0}$ and $y_{J}$ describing the JV lattice displacement.
The deformed PV stack with the coordinates
$\mathbf{u}_{n}=(u_{n,x},u_{n,y})$ generates the additional phase
difference
\[
\theta_{p}(\mathbf{r},\mathbf{u}_{n+1},\mathbf{u}_{n})=\arctan\frac
{y\!-\!u_{n+1,y}}{x\!-\!u_{n+1,x}}-\arctan\frac{y\!-\!u_{n,y}}{x\!-\!u_{n,x}}
\]
and modifies the Josephson energy. In addition to interactions
mediated by interlayer Josephson coupling, PVs are also aligned by
the magnetic interaction \cite{pancakes}. Due to its long-range
nature, this interaction can be well approximated by a simple
parabolic potential (see, e.g., Ref.\ \onlinecite{DodgsonPRL00})
with the magnetic-spring constant, $K_{p}\approx
s\Phi_{0}^{2}\mathcal{L} /\left( 4\pi\lambda^{2}\right)  ^{2}$ where
$\mathcal{L}\approx\ln (\lambda/r_{w})$ and
$r_{w}=\langle(u_{n+1}-u_{n})^{2}\rangle^{1/2}$ is the typical
wandering distance. Therefore in a homogeneous superconductor the
total energy change per one layer due to PV stack deformation is
given by the sum of the Josephson and magnetic energies
\begin{equation}
\delta\mathcal{E}=\frac{1}{N}\sum_{n}\left[  \left(  -1\right)  ^{n}
E_{J}\mathcal{C}(\mathbf{u}_{n+1},\mathbf{u}_{n},y_{J})+\frac{K_{p}}
{2}\mathbf{u}_{n}^{2}\right]  ,\label{JosEnTotal}
\end{equation}
where $N$ is the total number of layers and the function
$\mathcal{C} (\mathbf{u}_{2},\mathbf{u}_{1},y_{J})$ in the lowest
order with respect to the Josephson coupling energy, $E_{J}$, is
given by
\begin{align}
\mathcal{C}(\mathbf{u}_{2},\mathbf{u}_{1},y_{J})  &  =\int d^{2}\mathbf{r}
\left\{  \cos\left[  k_H(y-y_{J})\right]  \right.\nonumber\\
&\left.-\cos\left[ k_H(y-y_{J})+\theta
_{p}(\mathbf{r},\mathbf{u}_{2},\mathbf{u}_{1})\right] \right\}
\label{CDEF}\\
=\frac{I(k_H\mathbf{u}_{12})}{k_H^{2}}&\cos\left[  k_H\left(
y_{J}-\frac {u_{1,y}+u_{2,y}}{2}\right)  \right] \nonumber
\end{align}
with $\mathbf{u}_{12}\equiv\mathbf{u}_{2}-\mathbf{u}_{1}$ and
\begin{align}
I(\mathbf{v})  &  =\int dxdy\left[  \left(  1-\frac{r^{2}-v^{2}/4}{\left\vert
\mathbf{r}-\mathbf{v}/2\right\vert \left\vert \mathbf{r}+\mathbf{v}
/2\right\vert }\right)  \cos y\right. \nonumber\\
&  +\left.  \frac{v_{x}y-v_{y}x}{\left\vert \mathbf{r}-\mathbf{v}/2\right\vert
\left\vert \mathbf{r}+\mathbf{v}/2\right\vert }\sin y\right] \label{Idef}
\end{align}
with $\mathbf{v}\equiv k_H\mathbf{u}_{12}$. In the regime
$k_{H}|\mathbf{u}_{12}|\ll1$, this dimensionless function can be
computed analytically up to the third order with respect to the
reduced variable $\mathbf{v}$,
\begin{equation}
I(\mathbf{v})\!\approx\!\pi\!\left[  2v_{x}\!+\!\frac{v^{2}}{2}\left(
1\!-\!\frac{v_{x}} {4}\right)  \ln\left(  \frac{C_{h}}{v}\right)  \!
-\!\frac{v_{x}^{2}-v_{y}^{2}} {4}\!-\!\frac{v_{x}v_{y}^{2}}{8}\right]
\label{Iexpan}
\end{equation}
with $v^{2}=v_{x}^{2}+v_{y}^{2}\ $and $C_{h}=8\exp(-\gamma_{\mathrm
E})\approx4.492$ ($\gamma_{\mathrm E}\approx0.5772$ is the Euler
constant). The linear in $v_{x}$ term in $I(\mathbf{v})$ gives the
linear-displacement contribution to the Josephson energy, $-\left(
4\pi E_{J}/Nk_H\right)  \sum_{n}\left( -1\right)  ^{n}
u_{n,x}\cos\left[ k_Hy_{J}\right]  $. This term describes forces
acting on the straight PV stack from the alternating in-plane
currents induced by the JV lattice,
$j_{y}(y,n)\approx(-1)^{n}j_{h}\cos\left[  k_H(y-y_{J})\right]  $,
where $j_{h}=\left(  2\gamma/h\right)  j_{J}$, $h\equiv2\pi
B_{x}\gamma s^{2}/\Phi_{0}$ is the reduced magnetic field and
$j_{J}=(2\pi c/\Phi _{0})E_{J}$ is the Josephson current density.
Due to this term the ground state corresponds to the alternating PV
deformations along the direction of the in-plane field
\cite{Bul-zigzag}, $u_{n,x}=(-1)^{n}u_{a}$ with $u_{a}=\left(
s\Phi_{0}/c\right)  j_{h}/K_{p}=4\lambda^{2}/[h\gamma s\ln\left(
\lambda/u_{a}\right)  ]$, see Fig.\ \ref{Fig:JVL-PancStack}. For
such deformations, the quadratic term in the Josephson energy
cancels out and does not influence the deformation amplitude. The
assumed condition $k_Hu_{a}<1 $ is satisfied if $\gamma>\lambda/s$.

Lets consider now dynamic behavior. A transport current applied
across the layers drives the JV lattice along the layers. We
consider the lattice slowly moving with constant velocity through
the PV stacks along the $y$ axis, $y_{J}=vt$. Such motion generates
the electric field, $E_{z}=B_{x}v/c$. The lattice velocity is
assumed to be much smaller than the Swihart velocity so that the
lattice preserves its static structure. The homogeneously moving JV
lattice forces PVs to oscillate with the Josephson frequency,
$u_{x}(n,t)=(-1)^{n}u_{a}(t)$, with $u_{a}(t)=u_{a0}\cos(\omega_E
t+\alpha)$ and $\omega_E=k_Hv$. For a PV stack located near $y=0$,
the alternating deformation, $u_{a}(t)$, is described by a simple
oscillator equation
\begin{equation}
\eta_{p}\dot{u}_{a}+K_{p}u_{a}=f_{h}\cos\left(  \omega_E t\right)
,\label{PancDynam}
\end{equation}
where $f_{h}\equiv\left(  s\Phi_{0}/c\right)  j_{h}=4\pi\gamma sE_{J}/h$ and
$\eta_{p}$ is the pancake viscosity coefficient. Solution of this equation is
given by
\begin{equation}
u_{a}(t)=\operatorname{Re}\left[  \frac{f_{h}\exp(i\omega_E t)}{K_{p}
+i\omega_E\eta_{p}}\right].
\label{Solution}
\end{equation}
The frequency response of the stacks is determined by the relaxation
frequency, $\omega_{r}=K_{p}/\eta_{p}$. The oscillation amplitude
drops at $\omega_E >\omega_{r}$.

The average friction force per PV acting on the moving JV lattice is given by
\begin{widetext}
\begin{equation}
\mathcal{F}_{y}   =-\left\langle
\frac{d\delta\mathcal{E}}{dy_{J}}\right\rangle
=\frac{E_{J}}{Nk_H}\sum_{n}\left\langle \left(  -1\right)
^{n}I\left[ k_H\left(  \mathbf{u}_{n+1}\!-\!\mathbf{u}_{n}\right)
\right]  \sin\left[ \omega_E
t\!-\!k_H\frac{u_{n,y}\!+\!u_{n+1,y}}{2}\right]  \right\rangle
\label{FrictForce}
\end{equation}
\end{widetext}
where $\left\langle \ldots\right\rangle $ notates time
averaging. In the regime when only small alternating deformations in the
$x$ direction are present, this gives $\mathcal{F}_{y}=-4\pi
E_{J}\left\langle u_{a} (t)\sin\left(  \omega_E t\right)  \right\rangle $,
and, using the solution (\ref{Solution}), we obtain
\begin{equation}
\mathcal{F}_{y}=-2\pi
E_{J}f_{h}\frac{\omega_E\eta_{p}}{K_{p}^{2}+\omega_E^{2}
\eta_{p}^{2}},\label{PancFricForce}
\end{equation}
The average velocity of the JV lattice is connected with the applied
current, $j_{z}$, by the force-balance condition
\begin{equation}
-\eta_{\mathrm{Jff}}v+n_{p}\mathcal{F}_{y}+\left(  B_{x}/c\right)
j_{z} =0\label{ForceBalance}
\end{equation}
where $\eta_{\mathrm{Jff}}$ is the bare flux-flow friction
coefficient of the JV lattice and $n_{p}=B_{z}/s\Phi_{0}$ is the
concentration of PVs, and the third term is the driving force from
the transport current. This corresponds to the following
current-voltage dependence
\begin{equation}
j_{z}(E_{z})=\sigma_{\mathrm{Jff}}E_{z}+\delta
j(E_{z})\label{IVlowT}
\end{equation}
where $\sigma_{\mathrm{Jff}}=\left(  c/B_{x}\right)
^{2}\eta_{\mathrm{Jff}}$ is the bare flux-flow conductivity and
\begin{equation}
\delta j(E_{z})=-\frac{cn_{p}}{B_{x}}\mathcal{F}_{y}=\frac{\sigma_{p}E_{z}
}{1+\left(  sE_{z}/V_{r}\right)  ^{2}}\label{PancCurr}
\end{equation}
is the current enhancement due to the PV-induced damping of JV
motion, which we will call \emph{excess pancake-vortex current},
with $\sigma_{p}\equiv\sigma_{p}
(B_{z},B_{x})=2\pi^{2}n_{p}\eta_{p}s^{4}j_{h}^{2}/K_{p}^{2}$ and
$V_{r} =[\Phi_{0}/(2\pi c)]K_{p}/\eta_{p}$ is the voltage drop per
junction corresponding to the relaxation frequency. The
electric-field dependence of the conductivity $\delta
j(E_{z})/E_{z}$ resembles a well-known Drude frequency dependence.
Introducing the pancake flux-flow conductivity
$\sigma_{\mathrm{ff}}(B_{z})=\eta_{p}c^{2}/(s\Phi_{0}B_{z})$, we can
express the conductivity and voltage scales, $\sigma_{p}$ and
$V_{r}$, via experimentally-accessible parameters
\[
\sigma_{p}=2\sigma_{\mathrm{ff}}(B_{\lambda})\frac{B_{z}/B_{\lambda}}{\left(
\gamma h\right)  ^{2}},\ \
V_{r}=\frac{c\Phi_{0}}{8\pi^{2}\lambda^{2}\sigma
_{\mathrm{ff}}(B_{\lambda})}
\]
with $B_{\lambda}\equiv\Phi_{0}\mathcal{L}/(4\pi\lambda^{2})$. Note that
$\sigma_{p}$ scales with the field components as $\sigma_{p}\propto
B_{z}/B_{x}^{2}$.

\begin{figure}[ptb]
\begin{center}
\includegraphics[width=3.in]{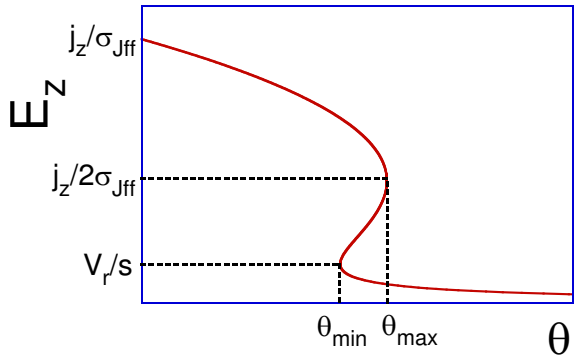}
\end{center}
\caption{The angular dependence of the electric field at fixed
current following from Eq.\ (\ref{Angle_Ez}) at $V_{r}\ll
sj_{z}/\sigma_{\mathrm{Jff}}$. Voltage jumps are expected at the
angles $\theta_{\max}$ and $\theta_{\min}$}
\label{Fig:Ez_thetaTheor}
\end{figure}

The key feature of the excess PV current, $\delta j(E_{z})$, is that
it depends nonmonotonically on the electric field. The maximum
current due to the pancake vortices can be estimated as
$j_{\max}=j_{J}\left( B_{z}/B_{\lambda }\right)  /h^{2}$. As
$\sigma_{p}\propto B_{z}$, the total I-V dependence (\ref{IVlowT})
becomes nonmonotonic at sufficiently large $B_{z}$. In fixed-current
experiments this leads to voltage jumps. Eq.\ (\ref{IVlowT}) also
determines the angular dependence of voltage at fixed current
frequently measured experimentally. Introducing the tilt angle of
field, $\theta \approx B_{z}/B_{x}\ll1$, we can rewrite Eqs.\
(\ref{IVlowT}) and (\ref{PancCurr}) in the form
\begin{equation}
\theta=h\left(  1+\left(  sE_{z}/V_{r}\right)  ^{2}\right)
\frac{j_{z} -\sigma_{\mathrm
{Jff}}E_{z}}{\sigma_{1}E_{z}}\label{Angle_Ez}
\end{equation}
with
$\sigma_{1}=4\sigma_{\mathrm{ff}}(B_{\lambda})\lambda^{2}/[\gamma^{3}
s^{2}\mathcal{L}]$. This equation explicitly determines the
dependence $E_{z}(\theta)$ at fixed $j_z$. In particular, in the
case $V_{r}\ll sj_{z}/\sigma_{\mathrm{Jff}}$, the maximum and
minimum of the dependence $\theta(E_{z})$ are given by
\begin{align*}
\theta_{\max}  & \approx \frac{h\left(  sj_{z}\right)  ^{2}}{8\sigma_{1}
\sigma_{\mathrm{Jff}}V_{r}^{2}}\ \text{at }E_{z}\approx j_{z}/2\sigma_{\mathrm{Jff}}\\
\theta_{\min}  & \approx\frac{2hsj_{z}}{\sigma_{1}V_{r}}\ \text{at }
E_{z}\approx V_{r}/s
\end{align*}
At these angles the voltage will have jumps. As we see, in a simple
model both angles are proportional to the in-plane field. The
angular dependence of the electric field following from Eq.\
(\ref{Angle_Ez}) at $V_{r}\ll sj_{z} /\sigma_{\mathrm{Jff}}$ is
illustrated in Fig.\ \ref{Fig:Ez_thetaTheor}.

\subsection{Friction force from fluctuating pancake-vortex stacks
\label{Sec:PVFluct}}

In real experimental conditions the result may be strongly
influenced by pinning and thermal fluctuations. Let us consider
influence of thermal fluctuations. At a finite temperature, in
addition to the regular zigzag deformations, the PV stack has random
fluctuating displacement $\mathbf{u} _{f,n}=(u_{f,x,n},u_{f,y,n})$.
These pancake fluctuations influence the damping of the JV lattice
in several different ways. The most direct channel is that, due to
random displacement in the $y$ direction (perpendicular to the
field), exact time periodicity of the force from the JV lattice
acting on PV is perturbed. This leads to the Debye-Waller-like
suppression of the pancake oscillation amplitude and reduction of
the corresponding current enhancement. The PV displacements also
modify the interaction force between PVs and JV lattice (in the low
orders with respect to displacements this effect is described by
Eq.\ (\ref{Iexpan})). Both above effects have the same order and
become strong when the typical PV fluctuation becomes of the order
of the JV lattice period. The pancake fluctuations also perturb
phase distribution of the JV lattice leading to renormalization of
its bare flux-flow resistivity. We will also estimate this effect.

We now analyze the fluctuation effects quantitatively. At finite
temperature and for the homogeneously moving JV lattice, $y_{J}=vt$,
the PV displacements obey the following equation
\begin{align}
&\eta_{p}\mathbf{\dot{u}}_{n}+K_{p}\mathbf{u}_{n}+\label{DynEq}\\
&(\!-\!1)^{n}E_{J}
\frac{\partial\left[\mathcal{C}(\mathbf{u}_{n\!+\!1},\mathbf{u}_{n},vt)
\!-\!\mathcal{C}(\mathbf{u}_{n} ,\mathbf{u} _{n\!-\!1},vt)\right]}
{\partial \mathbf{u}_{n}} \!=\!\mathbf{f} _{L,n}(t)\nonumber
\end{align}
where the function
$\mathcal{C}(\mathbf{u}_{1},\mathbf{u}_{2},y_{J})$ is defined by
Eqs.\ (\ref{CDEF}) and (\ref{Idef}) and $\mathbf{f}_{L,n}(t)$ is the
thermal Langevin force with the correlation function
\[
\left\langle
f_{L,n,\alpha}(t)f_{L,m,\beta}(t^{\prime})\right\rangle
=2\eta_{p}k_{\mathrm{B}}
T\delta_{nm}\delta_{\alpha\beta}\delta(t-t^{\prime}).
\]
This equation generalizes Eq.\ (\ref{PancDynam}) for finite
temperatures and large pancake displacements. Introducing the
dimensionless variables for the time, $\tilde{t} =\omega_{r}t$ and
coordinate $\mathbf{\tilde{u}}=k_{H}\mathbf{u}$, we conclude that
the overall behavior is determined by the two dimensionless
parameters,
\begin{equation}
\alpha_{J}=\frac{\pi E_{J}}{K_{p}}=\frac{\lambda^{2}}{(s\gamma)^{2}
\mathcal{L}}\label{alphaJ}
\end{equation}
giving the relative strength of magnetic and Josephson couplings
between PVs in different layers and the reduced temperature
\begin{equation}
\tilde{T}=\frac{k_H^{2}k_{\mathrm{B}}T}{K_{p}}=\frac{\left(  2\pi s\lambda
B_{x}\right) ^{2}}{\Phi_{0}^{2}\mathcal{L}}\frac{\left( 4\pi\lambda\right)
^{2}k_{\mathrm{B}} T}{s\Phi_{0}^{2}}.\label{redT}
\end{equation}
This reduced temperature has a transparent physical meaning, at
$\tilde{T}\sim 1$ the fluctuating pancake displacement becomes of
the order of the JV lattice period. We can see that the relative
strength of fluctuations rapidly increases with increasing in-plane
field.

\begin{figure}[ptb]
\begin{center}
\includegraphics[width=3.4in]{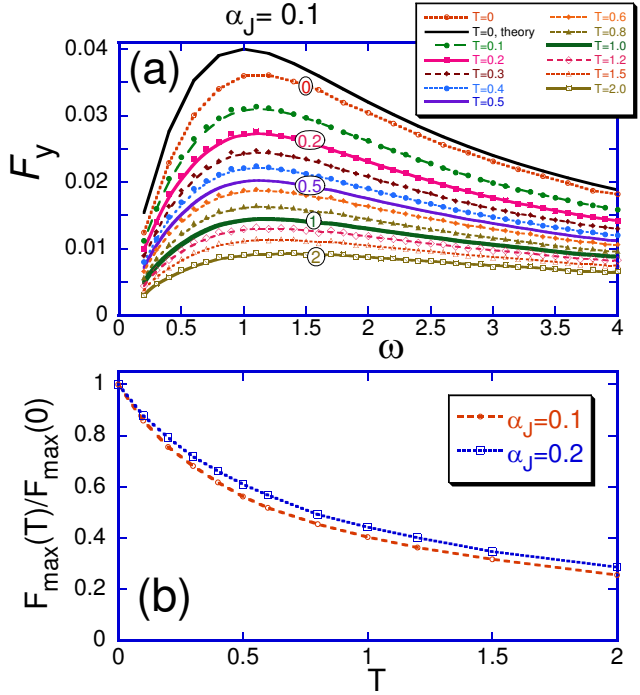}
\end{center}
\caption{(a)Evolution of the friction force vs Josephson frequency
dependence with increasing reduced temperature (\ref{redT}) for
$\alpha_{J}=0.1$. In this plot $\mathcal{F}_{y}$ is the friction
force induced by one fluctuating PV stack on JV lattice defined by
Eq.\ (\ref{FrictForce}) in units of $K_{p}/k_{H} $. The frequency
unit is $\omega_{r}$. The curves are calculated using pancake
displacements obtained by direct numerical simulation of the
Langevin dynamic equations (\ref{DynEq}). The upper solid line
represents the analytical result (\ref{PancFricForce}) valid at
$T=0$ and $\alpha_{J}\ll1$. (b)The temperature dependence of the
maximum friction force for two values of the parameter $\alpha_J$,
0.1 and 0.2.} \label{Fig:F_wE_num}
\end{figure}

We numerically solved the dynamic Langevin equations for the pancake
displacements $\mathbf{u} _{n}$ (\ref{DynEq}) with numerically
computed dependence $I(\mathbf{v})$ from Eq.\ (\ref{Idef}), and used
obtained solution to compute the average friction force
(\ref{FrictForce}). Figure \ref{Fig:F_wE_num}a shows the
Josephson-frequency dependences of the friction force
$\mathcal{F}_{y}$ for different reduced temperatures for typical
value of the parameter $\alpha_{J} $, $\alpha_{J}=0.1$. We can see
that the main effect of fluctuations is suppression of the maximum
of the $\mathcal{F}_{y}(\omega_E)$ dependence corresponding to the
maximum of $\delta j(E_{z})$. To quantify this suppression, we show
in Fig.\ \ref{Fig:F_wE_num}b temperature dependences of the maximum
friction force for two values of the parameter $\alpha_J$, 0.1 and
0.2. The maximum force decreases by factor of 2 at
$\tilde{T}\approx0.5-0.7$. The maximum only slightly shifts to
higher frequencies with increasing temperature. Also, at high
temperatures the force decays slower than $1/\omega_E$ at large
frequencies.

\subsection{Reduction of the regular flux-flow conductivity by
pancake-vortex fluctuations}

Pancake fluctuations also perturb regular phase distribution in the
moving JV lattice leading to modification of the regular flux-flow
conductivity. This gives an additional channel of PV influence on
the JV mobility. In this section we estimate this effect
quantitatively. Dynamic behavior of the JV lattice in layered
superconductors is described by the coupled reduced equations for
the interlayer phase differences $\theta_{n}$ and reduced fields
$h_{n}$, see, e.g., Ref.\ \onlinecite{KoshAranPRB01}
\begin{subequations}
\label{DynEqPhase}
\begin{align}
&\frac{\partial^{2}\theta_{n}}{\partial
t^{2}}\!+\!\nu_{c}\frac{\partial\theta_{n} }{\partial
t}\!+\!\sin\theta_{n}\!+\!\frac{\partial h_{y,n}}{\partial x}
-\frac{\partial h_{x,n}}{\partial y}   \!=\!0\label{EqPhase}\\
&\!\left(\!  \nabla_{n}^{2}\!-\!\frac{1}{l^{2}}\right)
h_{x,n}\!+\!\frac{\partial \theta_{n}}{\partial
y}\!+\!\nu_{ab}\frac{\partial}{\partial t}\left(
\frac{\partial\theta_{n}}{\partial
y}\!-\!\frac{h_{x,n}}{l^{2}}\right)\!=\!0 \label{EqField}
\end{align}
\end{subequations}
In these equations the units of space and time are given by the
Josephson length, $\lambda_{J}$, and the inverse plasma frequency,
$1/\omega_{p}$, $\nu_{c}$ and $\nu_{ab}$ are the dissipation
parameters related to the quasiparticle conductivities, and
$l=\lambda_{ab}/s$. For the moving triangular lattice the phase
differences can be represented as
\[
\theta_{n}(\mathbf{r},t)=\omega_{E}t-k_{H}y+\pi n+\theta_{v,n}(\mathbf{r}
,t)+\phi_{n}(\mathbf{r},t)
\]
where
$\theta_{v,n}(\mathbf{r},t)=\sum_{j}\theta_{p}(\mathbf{r}-\mathbf{R}
_{j},\mathbf{u}_{n+1,j}(t),\mathbf{u}_{n,j}(t))$ is the phase
perturbation induced by PV displacements, $\mathbf{u}_{n,j}(t)$, of
the stack located at $\mathbf{R}_{j}$ , $\phi_{n}(\mathbf{r},t)$ is
the correction due to the Josephson coupling. Solving Eqs.\
(\ref{DynEqPhase}) by Fourier transform in the first order with
respect to $\sin\theta_{n}$, we obtain
\begin{align}
\phi(\mathbf{k},q,\omega)  =&\frac{\left[  \sin\left(
\omega_{E}t\!-\!k_{H}y\!+\!\pi n\!+\!\theta_{v,n}(\mathbf{r},t)\right)
\right] _{\mathbf{k},\omega}}
{\Lambda(\mathbf{k},q,\omega)}\label{OscPhase}\\
\text{with~}\Lambda(\mathbf{k},q,\omega) &
=\!\omega^{2}\!-\!i\nu_{c}\omega
\!-\!\frac{k^{2}(1\!+\!i\nu_{ab}\omega)}{2(1\!-\!\cos q)\!+\!\left(
1\!+\!i\nu_{ab}\omega\right) /l^{2}},\nonumber
\end{align}
where $\left[  \ldots\right]  _{\mathbf{k},\omega}$ notates Fourier
transformation with respect to coordinate and time. To obtain I-V
characteristic we have to evaluate the average reduced Josephson
current $i_{J}=\left\langle \sin\theta_{n}\right\rangle $, which in
the first order with respect to $\phi_{n}(\mathbf{r},t)$ is given by
\[i_{J}\approx \left\langle \cos\left(  \omega_{E}t-k_{H}y+\pi
n+\theta_{v,n}(\mathbf{r},t)\right)
\phi_{n}(\mathbf{r},t)\right\rangle.\] Note that the PV excess
current derived in the previous sections also can be obtained using
this approach if we add to $\phi_{n}(\mathbf{r},t)$ the dynamic
phase perturbation due to the oscillating PVs.

Using result (\ref{OscPhase}), we derive
\begin{equation}
i_{J}=\int_{\mathbf{k},q,\omega}S(\mathbf{k},q,\omega)i_{J0}(\mathbf{k}
_{H}\!-\!\mathbf{k},\pi\!-\!q,\omega_{E}\!-\!\omega)\label{JosCurrGen}
\end{equation}
where $S(\mathbf{k},q,\omega)$ is the Fourier transform of the phase
correlation function
\begin{align*}
S_{n-n^{\prime}}(\mathbf{r}-\mathbf{r}^{\prime},t) &  =\left\langle
\exp\left[  i\left(  \theta_{v,n}(\mathbf{r},t)-\theta_{v,n^{\prime}
}(\mathbf{r}^{\prime},0)\right)  \right]  \right\rangle \\
&  \approx\exp\left[  \!-\!\left\langle \left(
\theta_{v,n}(\mathbf{r},t)
\!-\!\theta_{v,n^{\prime}}(\mathbf{r}^{\prime},0)\right)
^{2}\right\rangle\! /2\right]
\end{align*}
and
\[
i_{J0}(\mathbf{k},q,\omega)=\frac{1}{2}\operatorname{Im}\left[  \frac
{1}{\Lambda(\mathbf{k},q,\omega)}\right].
\]
In particular, without PV fluctuations
$S(\mathbf{k},q,\omega)=\delta (\mathbf{k})\delta(q)\delta(\omega)$
and the average Josephson current is simply given by
$i_{J0}(\mathbf{k}_{H},\pi,\omega_{E})$ giving at small $\omega_{E}$
the bare JV flux-flow resistivity (see Ref.\
\onlinecite{KoshAranPRB01}).

Let us evaluate the phase correlation function $S_{n}(\mathbf{r},t)$.
Neglecting correlations between displacements in different stacks, we
obtain
\begin{align}
&S_{n}(\mathbf{r}\!-\!\mathbf{r}^{\prime},t)=1\!-\!\frac{B_z}{\Phi_0}\!\int\!
d\mathbf{R}\!\left( 1\!-\!\exp\left[  -G(\mathbf{r}
\!-\!\mathbf{R},\mathbf{r} ^{\prime}\!-\!\mathbf{R}) \right]  \right)  \label{PancPhaseCorr}\\
&G(\mathbf{r} ,\mathbf{r} ^{\prime})\!=\!\frac{1}{2}\left\langle \left[
\theta_{p}(\mathbf{r},\mathbf{u}_{n+1}(t),\mathbf{u}_{n}(t))\!-\!\theta_{p}(\mathbf{r}
^{\prime},\mathbf{u}_{1}(0),\mathbf{u}_{0}(0))\right]
^{2}\right\rangle\nonumber
\end{align}
Using expansion of the PV phase difference at $r\gg |\mathbf{w}_n|$ with
$\mathbf{w}_n\equiv \mathbf{u}_{n+1}-\mathbf{u}_{n}$,
$\theta_{p}(\mathbf{r},\mathbf{u}_{n+1},\mathbf{u}_{n})\approx-\left[
\mathbf{r}\times \mathbf{w}_n \right]  _{z}/r^{2}$, we can see that the
integral over $\mathbf{R}$ logarithmically diverges at large $R$. Cutting
off this divergency at $R=R_{c}$, we obtain with logarithmical accuracy
the following result
\begin{equation}
S_{n}(\mathbf{r},t)\!=\!1\!-\!\pi n_{v}\left( \left\langle
r_{w}^{2}\right\rangle \ln\frac{R_{c}}{r_{w}}\!-\!\left\langle \mathbf{w}
_{n}(t)\mathbf{w}_{0}(0)\right\rangle \ln\frac{R_{c}}{|\mathbf{r}
|}\right) \label{PancPhCorrEst}
\end{equation}
with $\left\langle r_{w}^{2}\right\rangle \equiv\left\langle
\mathbf{w}_{n}^{2}\right\rangle $. From Eqs.\ (\ref{JosCurrGen}) and
(\ref{PancPhCorrEst}) we can conclude that the PV fluctuations
\emph{reduce} the regular contribution to the Josephson current and
the relative reduction can be estimated as $\delta
j_{r}/j\approx-\pi n_{v}\left\langle r_{w}^{2}\right\rangle
\ln\left( R_{c}/r_{w}\right) $. This correction roughly corresponds
to suppression of the Josephson coupling in the area $~\pi r_w^2$
around the PV stack.  At small velocities this reduction is
typically much smaller than the current \emph{increase} due to
pancake oscillations. However, as the PV excess current decays with
increasing velocity, the regular contribution may become dominating
at high velocities. The relative contribution of the regular term
also increases with increasing temperature.

\subsection{Role of columnar defects \label{Sec:ColDef}}

Columnar defects produced by heavy-ion irradiation are known to be
the most efficient pinners of PVs, see Ref.\ \onlinecite{PancCDpin}.
Moreover, they effectively suppress fluctuation and align the PV
stacks \cite{PancColDefPRB96}. Experimentally, this alignment leads
to significant increase of the Josephson Plasma Resonance frequency
\cite{CDsJPR}. Therefore we can expect that the columnar defects
will strongly suppress the PV-induced damping of JVs. Naively, one
may think that effect of the columnar defects can be described by a
simple enhancement of the effecting spring constant in the confining
parabolic potential. However, a more accurate analysis below shows
that this is incorrect description. A columnar defect produces a
very strong local potential for PVs, which is poorly described by
the parabolic approximation. Due to a discrete nature of the PV
stack, its statistical distribution around the columnar defect
consists of two parts \cite{PancColDefPRB96}: a sharp peak centered
at the defect corresponding to PV located inside the column and a
very wide envelope function corresponding to PV located outside the
column. In such situation, we can expect that PVs confined inside
the column do not contribute to JV damping while PVs outside the
column give almost the same contribution as free PVs. This means
that the columnar defect simply reduce the excess current by the
probability factor to find PV outside the columnar defect,
$P_{\mathrm{out}}$, as $\delta j\rightarrow P_{\mathrm{out}}\delta
j$. To estimate this probability, we consider a fluctuating pancake
vortex near the columnar defect, which we model as an insulating
disk with radius $b$. The ratio of probability to find the PV
outside the column, $P_{\mathrm{out}}$, to probability to find PV
inside the column, $P_{\mathrm{in}}$, can be evaluated as
\begin{equation}
\frac{P_{\mathrm{out}}}{P_{\mathrm{in}}}=\int\frac{d^{2}\mathbf{r}}{s_{0}}
\exp\left(  -\frac{\varepsilon_{v}(r)}{T}\right) \label{RelProbCD}
\end{equation}
where $\varepsilon_{v}(r)$ is the PV energy, measured with respect
to the ground state corresponding to the PV located in the column
center,
\[
\varepsilon_{v}(r)=U_{p}+s\varepsilon_{0}\ln\left(
1-b^{2}/r^{2}\right) +K_{p}r^{2}/2.
\]
with $\varepsilon_{0}\equiv (\Phi_0/4\pi \lambda)^2$. The first
term, $U_{p}\approx s\varepsilon_{0}\ln(b/\xi)$ is the pinning
energy for $b>\xi$, the second term is the interaction energy with
columnar defect, and the third term is the magnetic coupling energy,
$s_{0}$ is elemental area which can be evaluated by analyzing the
Gaussian fluctuation of the order parameter in the vicinity of
vortex core \cite{KoshelevPRB94},
\[
s_{0}\approx\frac{C_{s}2\pi\xi^{2}}{\ln(\kappa)}\frac{T}{s\varepsilon_{0}},
\]
in which $\xi$ is the coherence length, $\kappa=\lambda/\xi$ is the
Ginzburg-Landau parameter, and $C_{s} \sim1$.
In contrast to consideration of Ref.\ \onlinecite{PancColDefPRB96},
in the case of large in-plane field we can neglect Josephson
interaction in between PVs in neighboring layers and take into
account only magnetic coupling.
The value of the integral (\ref{RelProbCD}) is determined by the
competition between the two factors: the large energy cost
$s\varepsilon_{0}\ln(b/\xi)$ for putting PV outside the column and
large area of integration limited by the shallow magnetic-coupling
parabolic potential. For estimate, we can neglect the long-distance
tail $s\varepsilon_{0}\ln\left(  1-b^{2}/r^{2}\right)  $ in the
pinning potential. In this case we obtain the following estimate
\[
\frac{P_{\mathrm{out}}}{P_{\mathrm{in}}}\approx\exp\left(
-\frac{U_{p}} {T}\right)  \frac{2\pi T}{s_{0}K_{p}}\approx\exp\left(
-\frac{s\varepsilon _{0}\ln(b/\xi)}{T}\right)
\frac{\kappa^{2}}{C_{s}}
\]
Therefore, the probability to find PV outside the column can be
estimated as
\begin{equation}
P_{\mathrm{out}}\approx\frac{1}{1+\exp\left[  s\varepsilon_{0}\ln
(b/\xi)/T\right]  C_{s}/\kappa^{2}}\label{ProbResult}
\end{equation}
Ratio $P_{\mathrm{out}}/P_{\mathrm{in}}$ becomes of the order of $1$ at the
typical depinning temperature, $T^{\ast}$, which can be estimated as
\begin{equation}
T^{\ast}=\frac{s\varepsilon_{0}\ln(b/\xi)}{2\ln(\kappa)}\label{depinTemp}
\end{equation}
The columnar defects strongly reduce the damping due to PVs at $T\ll
T^{\ast} $ and this effect rapidly decreases at $T> T^{\ast}$. The
value of this important temperature scale is most sensitive to the
London penetration depth. For optimally doped BSCCO, assuming
$\lambda(T)=200$nm$/\sqrt(1-(T/T_c)^2)$, $\xi
=15$nm$/\sqrt(1-(T/T_c)^2)$ with $T_c=90$K, and $b=7$nm, we estimate
$T^{\ast}\approx 58$K. For weakly underdoped BSCCO, taking
$\lambda(0)=250$nm and $T_c=85$K, we obtain $T^{\ast}\approx 45$K.

\section{Experimental results \label{Sec:Experiment}}

\begin{figure}[ptb]
\begin{center}
\includegraphics[width=3.in]{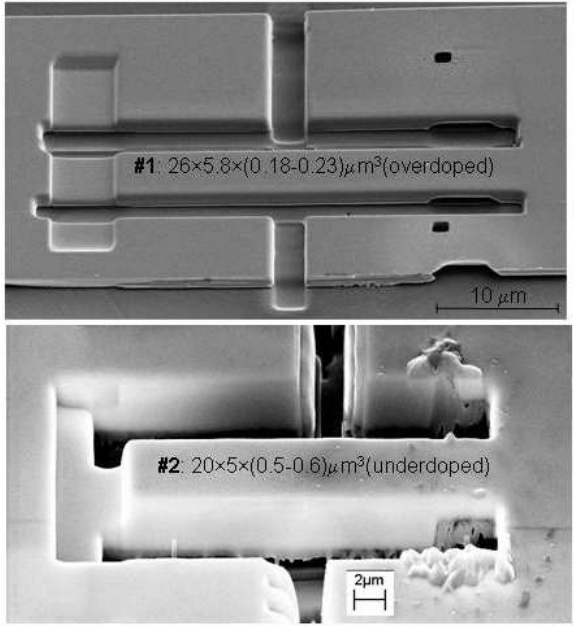}
\end{center}
\caption{Two mesa structures fabricated out of BSCCO whiskers for
transport measurements.} \label{Fig:mesas}
\end{figure}

Josephson flux-flow (JFF) measurements have been done on the two
mesa structures fabricated from the BSCCO single-crystal whiskers
using double-sided processing by the focused ion beam (FIB)
technique \cite{FIB}, see Fig.\ \ref{Fig:mesas}. The mesas had
geometrical sizes L$_{a}\times$L$_{b}\times$L$_{c}$=26$\mu$m$\times
$5.8$\mu$m$\times$(0.18-0.23)$\mu$m (mesa \#1) and
20$\mu$m$\times$5$\mu $m$\times$(0.5-0.6)$\mu$m (mesa \#2). The mesa
\#1 was prepared out of the overdoped whisker grown in oxygen flow.
The mesa \#2 was prepared out of the underdoped whisker grown in a
mixture of 10-20\ of O$_{2}$ and 80-90\% of Ar.

At fixed temperatures and fixed in-plane component of the magnetic
field ($H\parallel b$) we measured the JV flux-flow resistance
$R_{\mathrm{ff}}$ at several fixed currents and the I-V
characteristics as a function of the c-axis component of the field
($H\parallel$ c). The cryostats with two perpendicular coils have
been used for these experiments, the main coil provided the parallel
field component, $H_{x}$, while another coil induced the
perpendicular component, $H_{z}$. The strictly parallel orientation
has been found using sharp maximum of $R_{\mathrm{ff}}(H_{z})$
dependence by sweeping $H_{z}$.
We estimated an accuracy of field rotation in this way to be within
0.01$^{\circ}$. The experiments have been carried out at high
temperatures above 40K to avoid flux trapping effects. As a separate
part of the experiments, we studied the influence of columnar
defects on the JFF with and without the perpendicular field
component. The irradiation was made by Pb-ions with energy of 1 GeV
at the GANIL accelerator (Caen, France). The defects were introduced
into the mesa \#2 along the \textit{c} axis with concentration $3
\cdot 10^{8}$cm$^{-2}$ corresponding to the matching field of about
60 Oe. For a comparison, the JFF characteristics have been measured
in the mesa \#2 first just before irradiation and then remeasured
exactly at the same conditions within a week after irradiation.

\begin{figure}[ptb]
\begin{center}
\includegraphics[width=3.4in]{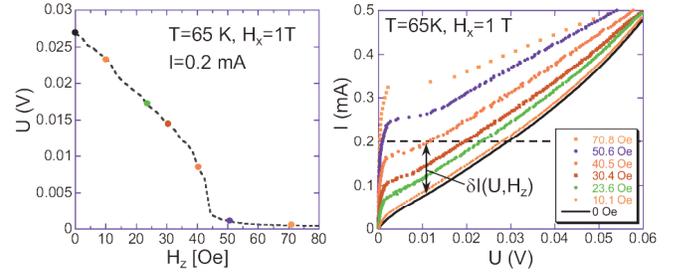}
\end{center}
\caption{ Left plot shows $H_{z}$ dependence of the z-axis voltage
at fixed current $I=$0.2mA and in-plane filed $H_{x}=1T$ for the
mesa \#1. The right plot shows the current-voltage (I-V) dependences
at several fixed $H_{z}$ values marked in the left plot. The
horizontal dashed line marks the current value in the left plot. One
can see that the I-V dependence acquires a jump around
$H_{z}\approx50 $Oe, which is a coming from the nonmonotonic
contribution, $\delta I(U,H_{z})$, from the pancake vortices. As a
consequence, the $H_{z}$ (or angular) dependence of voltage at fixed
current in the left side also has a jump. } \label{Fig:AngScanIVs}
\end{figure}

\begin{figure}[ptb]
\begin{center}
\includegraphics[width=3.2in]{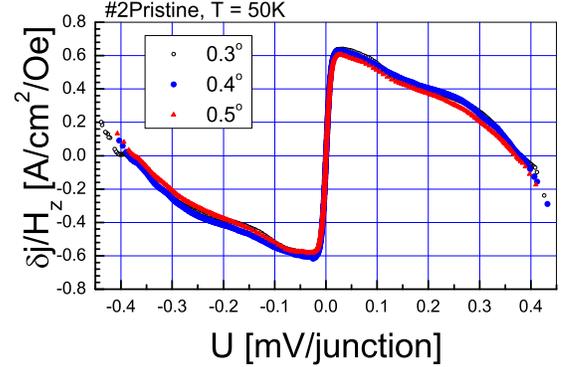}
\end{center}
\caption{Plots of the ratio $\delta j/H_{z}$ vs voltage for
$H_{x}=1T $ and three tilt angles demonstrating that the excess
pancake current is proportional to concentration of PVs.}
\label{Fig:djHzScaling}
\end{figure}
\begin{figure}[ptb]
\begin{center}
\includegraphics[width=3.2in]{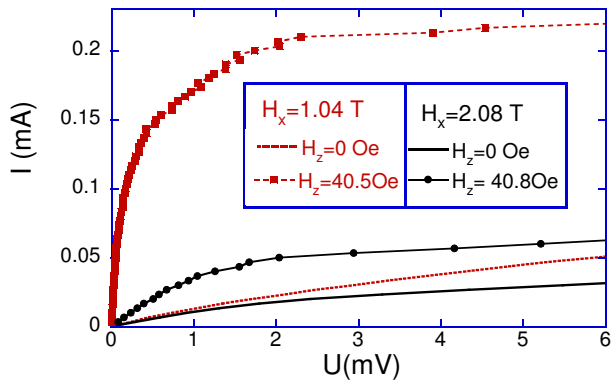}
\end{center}
\caption{The current-voltage dependences of the mesa \#1 for two
values of the in-plane field, $H_x=1.04$T and $2.08$T, and for very
close values of c-axis field $H_z=40.5$ and $40.8$Oe. For reference,
the I-V dependences at $H_z=0$Oe are also shown for both $H_x$. The
amplitude of the excess PV current for $H_x=2.08$T is more than four
times smaller than for $H_x=1.04$T.} \label{Fig:Mesa1Hz40_5Oe}
\end{figure}
\begin{figure}[ptb]
\begin{center}
\includegraphics[width=3.2in]{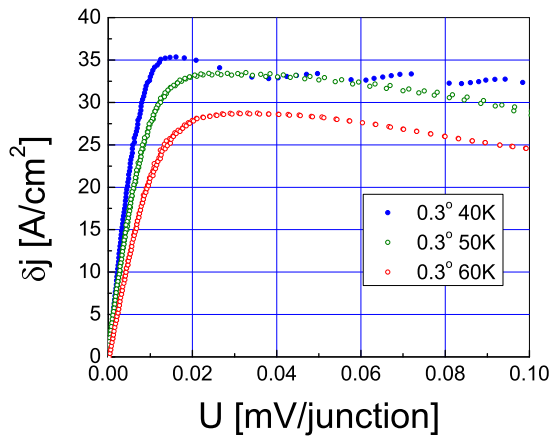}
\end{center}
\caption{Temperature evolution of the excess PV current density for
the mesa \#2 at $H_{x}=1$T and tilt angle 0.3$^{\circ}$
corresponding to $H_{z}=52.4$Oe. The voltage corresponding to
maximum excess current decreases with decreasing temperature.}
\label{Fig:IVsTempCompare}
\end{figure}
\begin{figure}[ptbptb]
\begin{center}
\includegraphics[width=2.8in]{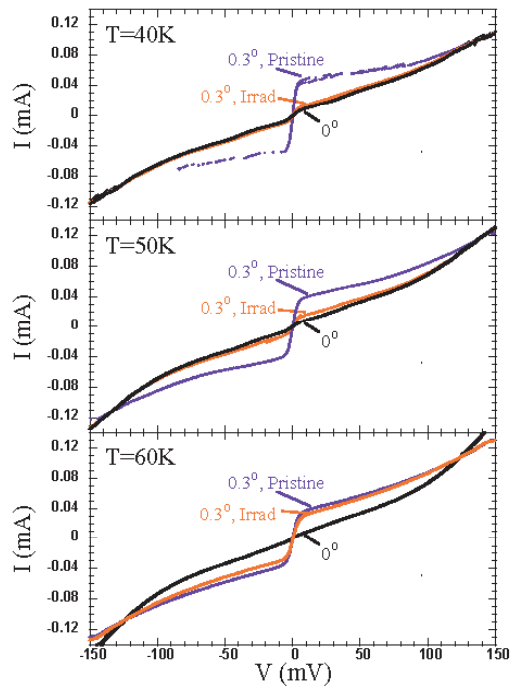}
\end{center}
\caption{Influence of heavy-ion irradiation on the pancake excess
current. The current-voltage characteristics are shown for mesa \#2
before and after irradiation, for the magnetic field $1T$ tilted at
angle $0^{\circ}$ and $0.3^{\circ} $ (corresponding to $H_{z}\approx
52$Oe). Irradiation dose corresponds to the matching field,
$B_{\Phi}=60$G. Irradiation does not influence motion of the JV
lattice without PVs, at $H_{z}=0$ but strongly reduces pancake
effect at $T=40$K and $50$K. This reduction is much weaker at
$60$K.} \label{Fig:dIComp03degTemp}
\end{figure}

Figure \ref{Fig:AngScanIVs}(left panel) shows the dependence of the
JFF voltage at fixed current on the \textit{c}-axis magnetic field,
$H_{z}$ for the mesa \#1. We can see that the JFF voltage is highly
sensitive to a small component of $H_{z}$ resulting in large
negative JFF magnetoresistance $R_{JFF}(H_{z})=U(H_{z})/I$. At the
fixed current (0.2 mA) and fixed in-plane field (1 tesla) the JFF
resistance first decreases linearly with $H_{z}$ and then drops down
sharply to the low-resistance state. To understand the mechanism of
this drop, we show in the right panel of Fig.\ \ref{Fig:AngScanIVs}
a series of I-V dependences traced at the different fixed $H_{z}$
values marked in the left plot. We can see from this plot, that the
voltage drop to the low-resistance state originates from a
nonmonotonic character of the I-V dependence which appears at high
enough $H_{z}$. The excess current, $\delta I(U,H_{z})\equiv
I(U,H_{z})-I(U,0)$, in the I-V dependences is a direct consequence
of an additional damping of the moving JV lattice induced by the PV
stacks, as discussed in the theoretical sections above. Figure
\ref{Fig:djHzScaling} shows plots of the ratios $\delta j/H_{z}$ vs
bias voltage $U$ for the mesa \#2 before irradiation at different
tilt angles corresponding to different $H_{z}$. One can see
nonmonotonic dependence $\delta j(U)$, that first increases rapidly,
reaches a maximum and then decreases slowly. Another remarkable
feature is proportionality of $\delta j$ to $H_{z}$ up to
$H_{z}\approx 87$ Oe (0.5$^{\circ}$ tilt), which is seen from an
almost ideal collapse of the $\delta j/H_{z}$ curves at different
tilt angles. Figure \ref{Fig:Mesa1Hz40_5Oe} shows the I-V
dependences of the mesa \#1 for two values of the in-plane field,
$H_x=1.04$T and $2.08$T, and for very close values of c-axis field,
$H_z\approx 40.5$Oe. We can see that for $H_x=2.08$T the amplitude
of the excess current is more than four times smaller than for
$H_x=1.04$T, which is consistent with the theoretical estimates.
Figure \ref{Fig:IVsTempCompare} illustrates the temperature
dependence of the excess current for the mesa \#2. We can see that
at lower temperatures maximum in the excess current increases and
moves to the lower voltages. We also note that for the overdoped
mesa the excess PV current density is 4-5 times larger than for the
underdoped one.

Figure \ref{Fig:dIComp03degTemp} shows comparison of the PV-induced
damping for the mesa \#2 before and after irradiation for three
temperatures, 40K, 50K, and 60K. It is interesting to note that
\emph{the columnar defects induced by heavy-ion irradiation have
negligible direct influence on dynamics of JV lattice} (IVs at
$0^{\circ}$ tilt angles are not noticeably influenced by the
irradiation). On the other hand, they strongly influence the
PV-induced damping at low temperatures. We can see that at 40K and
50K pinning of the PV stacks by the columnar defects almost
completely eliminates the PV extra damping. At 60 K this damping is
restored most probably due to the thermal depinning of the PV stacks
from the columns. This behavior is consistent with location of the
thermal depinning temperature $T^{\ast}$ evaluated in Sec.\
\ref{Sec:ColDef} (for underdoped BSCCO we estimated $T^{\ast}\sim
45$K).

\section{Discussion \label{Sec:Discussion}}

Strong dynamic interaction of the JVs with the PV stacks leads to a
very high sensitivity of the JV transport to a very small
concentration of the PV stacks which we observed experimentally. At
a given driving force (or DC current) that interaction slows down
the JV lattice and thus reduces the JFF voltage leading to the
pronounced negative JFF magnetoresistance with respect to the
\textit{c}-axis component of the magnetic field. On the other hand,
at a given voltage the extra damping due to the PV stacks results in
increase of the current. Theoretical description of these effects is
based on calculation of the oscillating zigzag deformations of the
PV stacks induced by the moving JV lattice and the average friction
force due to these deformations. Several features of the
experimental PV excess current, such as its nonmonotonic dependence
on voltage, its decrease with increasing temperature and the
in-plane field, are in a good \emph{qualitative} agreement with the
theory. A strong suppression of the PV damping by columnar defects
also provides confirmation for suggested mechanism. A detailed
\emph{quantitative} comparison between the experiment and theory
requires an accurate knowledge of the microscopic parameters of
superconductor, London penetration depth, $\lambda_{ab}$, anisotropy
$\gamma$, and pancake flux-flow resistivity $\rho_{\mathrm{ff}}$, as
well as geometrical parameters of the mesas.

To estimate the key theoretical parameters $\sigma_{p}$ and $V_{r}$
we need values of the magnetic spring constant $K_{p}$ and the PV
viscosity coefficient, $\eta_{p}$. The magnetic spring constant is
mainly determined by the London penetration depth. For example,
taking a value $\lambda=300$nm typical for overdoped BSCCO at
$T=65$K, we obtain $K_{p}/s\approx 65$N/m$^{2}$ and
$B_{\lambda}\approx 36$G. The value of the viscosity coefficient is
more uncertain. To our knowledge no direct transport measurement of
the PV flux-flow resistivity in BSCCO has been published. Microwave
measurements of the flux-flow resistivity in single crystals (at
40.8 GHz) \cite{HanaguriPRL99} and films (at 48 GHz)
\cite{SilvaSST00} show that $\eta_p\propto
d\rho_{\mathrm{ff}}/dB_{z}$ has strong temperature dependence and
suggest a typical value for $T=60$K,
$\eta_p/s$$\approx$$10^{-8}$-$10^{-7}$N s/m$^{2}$ corresponding to
the slope $d\rho _{\mathrm{ff}}/dB_{z}=2$-$20$ $\mu\Omega$ cm/tesla.
This gives the following estimate for the flux flow-conductivity at
$B_{z}=B_{\lambda}$,
$\sigma_{\mathrm{ff}}(B_{\lambda})$$\sim$$2\cdot10^{8}$-$2\cdot10^{9}$[$\Omega$\
cm]$^{-1}$. These data suggest that the maximum PV effect is
expected at very small voltage, $V_{r}\sim0.1$-$1\mu$V.
Experimentally, the maximum excess current is observed at
significantly higher voltages, $V_{r}\sim 15$-$30\mu$V corresponding
to relaxation frequencies 7-15 GHz. Also, the simple theory does not
quite describe shapes of the experimental $\delta j (U)$
dependences, at $U>V_{r}$ the excess current decreases slower than
expected $1/U$ decay. On the other hand, we found that shapes of
$\delta j(U)$ dependences also somewhat differ for underdoped and
overdoped mesas. These discrepancies can be explained if we
(i)assume small-velocity value for the viscosity constant, $\eta_p/s
\sim 2\cdot 10^{-6}$N s/m$^2$, which is significantly larger than
the microwave measurements suggest, and (ii)assume that it decreases
with increasing PV velocity. In the temperature and in-plane field
region probed in experiment influence of thermal fluctuation PV's
considered in Sec. \ref{Sec:PVFluct} is significant. For example,
for the overdoped mesa at 65 K, we estimate $\alpha_J\approx
0.15$-$0.2$ and the reduced temperature given by Eq.\ (\ref{redT})
as $\tilde{T}\approx 0.2$ for $B_x=1$T and $\approx 0.8$ for
$B_x=2$T. Shape deviations, at least partly, can be due to the PV
thermal fluctuations. Our calculations show that the PV fluctuations
mainly reduce the amplitude of $\delta j$ and somewhat weaken the
dependence $\delta j(U)$ at large $U$ but they do not shift much the
maximum of the $\delta j(U)$ dependence.

Another possible reason for disagreement with the simple theoretical
picture is that at small velocities the JV lattice may move
inhomogeneously, via thermally-activated jumps. This regime of
motion is not considered by the theory at all and it may lead to an
increase of apparent $V_r$. We expect that this inhomogeneous regime
is more pronounced for smaller sample sizes in which averaging over
PV locations is incomplete. Therefore, a systematic size dependence
of the effect would help to understand mechanism of the low-voltage
regime.

Note that, due to a very small value of $V_{r}$, without
scrutinizing I-V shape, the low-voltage enhancement of damping can
be easily misinterpreted as an enhancement of the critical current.
In our experiments, a detailed study of the small-voltage region
shows that the JV lattice slowly moves below the maximum current
meaning that this maximum is due to the enhancement of damping.
This, of course, does not exclude possibility that in other
experimental conditions (e.g., at low temperatures) the main PV
effect would be enhancement of the critical current.

An extreme sensitivity of the JV flux-flow voltage to a very small
concentration of the PVs suggests a very attractive possibility to
use small-size mesas for detection of penetration of individual PV
stacks which will be seen as current or voltage jumps. The change of
the total current due to the penetration event of a single PV stack,
$\delta I_{1}$, is expected to be independent on the mesa size and
is given by
\[
\delta I_{1}=\frac{d\delta j}{dB_{z}}\Phi_{0}.
\]
Its maximum value, $\delta I_{1,\max}$, is realized at $V=V_r$ and
at low temperatures, where thermal-fluctuation suppression is small,
can be estimated as
\[
\delta I_{1,\max}=j_{J}\frac{4\pi\lambda^{2}}{h^{2}\ln(\lambda/r_{w})}.
\]
Taking $j_{J}=500$A/cm$^{2}$, $\lambda= 300$nm, and $h=2$, we
estimate $\delta I_{1,\max}\approx0.5\mu$A. On the other hand, for
the underdoped mesa \#2 at 50K from Fig.\ \ref{Fig:djHzScaling} we
obtain $\delta I_{1,\max}\approx0.12\mu$A consistent with the above
theoretical estimate. For the overdoped mesa $\delta I_{1,\max}$ is
4-5 times larger. Such currents are detectable experimentally. We
conclude that this technique can be practically used for detection
of the individual PV stacks penetration.

\section{Acknowledgements}

We would like to thank A.\ M.\ Nikitina for providing the BSCCO
single-crystal whiskers, A.\ Ryhd, G.\ Karapetrov, and M.\ Gaifullin
for valuable technical assistance. AEK also would like to thank K.\
Kadowaki and Y.\ Kakeya for useful discussions. Work in Argonne was
supported by the U.S. DOE, Office of Science, under contract \#
W-31-109-ENG-38. AEK and YIL acknowledge support from the NATO
Travel grant No.\ PST.CLG.979047. The work in Ecole Polytechique was
partially performed in the frame of the CNRS-RAS Associated European
Laboratory between CRTBT and IRE "Physical properties of coherent
electronic states in condensed matter".


\begin{thebibliography}{99}
\bibitem{pancakes}
K.B. Efetov, Zh. Eksp. Teor. Fiz. \textbf{76}, 1781 (1979) [Sov.
Phys. JETP \textbf{49}, 905 (1979)]; A.\ I.\ Buzdin and D.\
Feinberg, {\em J. Phys. (Paris)} \textbf{51}, 1971 (1990);
%
S.\ N.\ Artemenko and A.\ N.\ Kruglov, Phys.\ Lett.\ A \textbf{143}
485 (1990); J.\ R.\ Clem, Phys.\ Rev.\ B \textbf{43}, 7837 (1991).
\bibitem{BulClemPRB91}L.\ N.\ Bulaevskii and J.\ R.\ Clem, Phys.\ Rev.\ B,
\textbf{44}, 10234 (1991).
\bibitem{BulLedvKoganPRB92}L.~N.~Bulaevskii, M.~Ledvij, and V.~G.~Kogan,
Phys.~Rev.\ B \textbf{46}, 366 (1992).

\bibitem{CrossLatLet}A.\ E.\ Koshelev, Phys.\ Rev.\ Lett. \textbf{83}, 187
(1999).
\bibitem{Bolle91}C.\ A.\ Bolle, P.\ L.\ Gammel, D.\ G.\ Grier,
C.\ A.\ Murray, D.\ J.\ Bishop, D. B. Mitzi and A. Kapitulnik,
Phys.\ Rev.\ Lett.\ \textbf{66}, 112 (1991).
\bibitem{MatsudaSci02}T.\ Matsuda, O.\ Kamimura, H.\ Kasai, K.\ Harada,
T.\ Yoshida, T.\ Akashi, A.\ Tonomura, Y.\ Nakayama, J.\ Shimoyama,
K.\ Kishio, T.\ Hanaguri, and K.\ Kitazawa, Science, \textbf{294},
2136(2001).

\bibitem{Grig95}I.\ V.\ Grigorieva, J.\ W.\ Steeds, G.\ Balakrishnan, and
D.\ M.\ Paul, Phys.\ Rev.\ B \textbf{51}, 3765 (1995).
\bibitem{GrigNat01}A.\ Grigorenko, S.\ Bending, T.\ Tamegai, S.\ Ooi, and
M.\ Henini, Nature \textbf{414}, 728 (2001).
\bibitem{VlaskoPRB02}V.\ K.\ Vlasko-Vlasov, A.\ E.\ Koshelev, U.\ Welp,
G.\ W.\ Crabtree, and K.\ Kadowaki, Phys.\ Rev.\ B \textbf{66}, 014523 (2002).

\bibitem{TokunagaPRB02}M. Tokunaga, M. Kobayashi, Y. Tokunaga, and T.
Tamegai, Phys.\ Rev.\ B \textbf{66}, 060507(R) (2002).
\bibitem{ChainReview}S.\ Bending and M. J. W. Dodgson, J. Phys.: Condens.
Matter \textbf{17}, R955 (2005).
\bibitem{Lee}J.\ U.\ Lee, J.\ E.\ Nordman, and G.\ Hohenwarter,
Appl.\ Phys.\ Lett., \textbf{67}, 1471 (1995); J.\ U.\ Lee , P.\ Guptasarma,
D.\ Hornbaker, A.\ El-Kortas, D.\ Hinks, and K.\ E.\ Gray Appl.\ Phys.\ Lett.,
\textbf{71}, 1412 (1997).

\bibitem{Hechtfischer}G.\ Hechtfischer, R.\ Kleiner, A.\ V.\ Ustinov, and
P.\ M\"{u}ller, Phys.\ Rev.\ Lett.\ \textbf{79}, 1365 (1997);
G.\ Hechtfischer, R.\ Kleiner, K.\ Schlenga, W.\ Walkenhorst, P.\ M\"{u}ller,
and H.\ L.\ Johnson, Phys.\ Rev.\ B, \textbf{55}, 14638 (1997).

\bibitem{Latyshev}Yu.\ I.\ Latyshev, P.\ Monceau, and V.\ N.\ Pavlenko,
Physica C \textbf{282-287}, 387 (1997); Physica C \textbf{293}, 174 (1997);
Yu.\ I.\ Latyshev, M.\ B.\ Gaifullin, T.\ Yamashita, M.\ Machida, and
Y.\ Matsuda, Phys.\ Rev.\ Lett., \textbf{87}, 247007 (2001).
\bibitem{Bul-zigzag}L.\ N.\ Bulaevskii , M.\ Maley, H.\ Safar, and
D.\ Dom\'{\i}nguez, Phys.\ Rev.\ B \textbf{53}, 6634 (1996).

\bibitem{LatPhysC91}Yu.\ Latyshev and A.\ Volkov, Physica C, \textbf{182}, 47 (1991)

\bibitem{EnriquezPRB96}H.\ Enriquez, N.\ Bontemps, P.\ Fournier,
A.\ Kapitulnik, A.\ Maignan, and A.\ Ruyter Phys.\ Rev.\ B, \textbf{53},
R14757 (1996)

\bibitem{HechtfischerPRB97}G.\ Hechtfischer, R.\ Kleiner, K.\ Schlenga,
W.\ Walkenhorst, P.\ M\"{u}ller, and H.\ L.\ Johnson Phys.\ Rev.\ B
\textbf{55}, 14638 (1997).

\bibitem{JVPancPRB03}A.\ E.\ Koshelev, Phys.\ Rev.\ B \textbf{68}, 094520
(2003).


\bibitem{DodgsonPRL00}M. J. W. Dodgson, A. E. Koshelev, V. B. Geshkenbein,
and G. Blatter, Phys.Rev.Lett. \textbf{84}, 2698 (2000).

\bibitem{KoshAranPRB01}A.\ E.\ Koshelev and I. Aranson,
Phys.\ Rev.\ B \textbf{64}, 174508 (2001).

\bibitem{PancCDpin}L.~Civale, A.~D. Marwick, T.~K. Worthington, M.~A. Kirk, J.~R. Thompson,
L.~Krusin-Elbaum, Y.~Sun, J.~R. Clem, and F.~Holtzberg, Phys. Rev.
Lett., {\bf 67}(5), 648 (1991);
%
M.~Konczykowski, F.~Rullier-Albenque, E.~R. Yacoby, A.~Shaulov,
Y.~Yeshurun,  and P.~Lejay, Phys. Rev. B, {\bf 44}, 7167 (1991);
%
W.~Gerh\"auser, G.~Ries, H.~W. Neum\"uller, W.~Schmidt, O.~Eibl,
G.~Saemann-Ischenko, and S.~Klaum\"unzer, Phys. Rev. Lett., {\bf
68},  879 (1992);
%
L. Civale, Supercond. Sci. Technol. \textbf{10} A11 (1997);
%
C. J. van der Beek, M. Konczykowski, R. J. Drost, P. H. Kes, N.
Chikumoto, and S. Bouffard, Phys. Rev. B \textbf{61}, 4259 (2000).

\bibitem{PancColDefPRB96}A.E.Koshelev, P. Le Doussal, and V.M.Vinokur,
Phys.\ Rev. B \textbf{53} R8855 (1996).

\bibitem{CDsJPR}M. Sato, T. Shibauchi, S. Ooi, T. Tamegai, and M.
Konczykowski,  Phys.\ Rev.\ Lett. \textbf{79}, 3759 (1997); M.
Kosugi, Y. Matsuda, M. B. Gaifullin, L. N. Bulaevskii, N. Chikumoto,
M. Konczykowski, J. Shimoyama, K. Kishio, K. Hirata, and K. Kumagai,
Phys. Rev. Lett. \textbf{79}, 3763 (1997)

\bibitem{KoshelevPRB94}A.E.Koshelev, Phys.Rev. \textbf{50} 506 (1994).

\bibitem{FIB}Yu.\ I.\ Latyshev, S.-J. Kim, and T. Yamashita, IEEE
Trans. on Appl. Supercond., 9, 4312(1999).

\bibitem{HanaguriPRL99}T. Hanaguri, T. Tsuboi, Y. Tsuchiya, Ken-ichi Sasaki,
and A. Maeda, Phys.\ Rev.\ Lett. \textbf{82}, 1273 (1999).

\bibitem{SilvaSST00}E. Silva, R. Fastampa, M. Giura, R. Marcon, D. Neri, and
S. Sarti, Supercond. Sci. Technol., \textbf{13} 1186 (2000).
\end{thebibliography}
\end{document}